\documentclass[prc,twocolumn]{revtex4}

\usepackage{graphicx}
\usepackage{color}
\usepackage{bm}
\usepackage{amsmath, amssymb}
\usepackage{type1cm}
\usepackage{comment}
\usepackage{url}

\begin{document}

\title{ $SO(3)$ quadratures in angular-momentum projection }

\author{
  Noritaka Shimizu$^{1,2}$\footnote{shimizu@nucl.ph.tsukuba.ac.jp},
  and 
  Yusuke Tsunoda$^1$}
\affiliation{
  $^1$Center for Computational Sciences,
  University of Tsukuba, 1-1-1 Tennodai, Tsukuba, Ibaraki 305-8577, Japan\\
  $^2$Center for Nuclear Study, The University of Tokyo, 
  7-3-1 Hongo, Bunkyo-ku, Tokyo 113-0033, Japan
}

\begin{abstract}
  While the angular-momentum projection is a common tool for theoretical nuclear structure studies, a large amount of computations are required particularly for triaxially deformed states. 
  In the present work, we clarify the conditions of the exactness of quadratures in the projection method. For efficient computation, the Lebedev quadrature and spherical $t$-design are introduced to the angular-momentum projection.
  The accuracy of the quadratures is discussed in comparison with the conventional Gauss-Legendre and trapezoidal quadratures. We found that the Lebedev quadrature is the most efficient among them and the necessary number of sampling points for the quadrature, which is often proportional to the computation time, is reduced by a factor 3/2 in comparison with the conventional method.
\end{abstract}

\maketitle

\section{Introduction}
\label{sec:intro}

% Historical ... mention Gaussian Overlap approximation? 

The angular-momentum projection has been widely used and is a crucial technique for nuclear structure calculations \cite{ringschuck}.
A mean-field method often provides us with a spontaneously symmetry-broken wave function, whose symmetry can be restored by the projection method. 
The projection method plays an important role in various mean-field and beyond-mean-field approaches such as the projected shell model \cite{psm}, 
the VAMPIR method \cite{vampire},
the projected configuration interaction method \cite{pci},
the Fermionic molecular dynamics \cite{fmd_npa2004},
the Monte Carlo shell model (MCSM) \cite{phys_scr_mcsm}, 
the quasi-particle vacua shell model \cite{qvsm},
the hybrid multi-determinant method \cite{puddu}, 
the projected variational Monte Carlo method \cite{vmcsm},
% the shell model Monte Carlo method \cite{smmc}, 
and the generator coordinate method (GCM) with 
angular-momentum projection (e.g. 
 \cite{rpgcm,rodriguez_gcm,amd_proj}).
The angular-momentum projection is not only a tool to obtain an accurate wave function but also a connection between the intrinsic wave function and that in the laboratory frame \cite{ringschuck,tplot}.

The case of the axially symmetric deformation requires only a one-dimensional integral for the angular-momentum projection.
However, to include the degree of freedom of triaxial deformation, the projection demands a three-fold integral of the Euler angles and is computationally expensive.
Recently developed configuration-mixing methods and variation-after-projection methods require accurate numerical evaluation of the projected matrix elements.
Since the computational resource for the projection procedure is approximately proportional to the number of the sampling points in the quadrature to evaluate the integral of Euler angles in the projection operator, it is greatly valuable to find an efficient way to reduce the number of the points as much as possible and to save the computation time in various theoretical frameworks containing the angular-momentum projection. Conventionally, a product of the trapezoidal and the Gauss-Legendre quadratures has been widely used for this purpose.
A comprehensive discussion of the projection method is found in Ref.~\cite{bally_proj}. Recently, a projection method using linear algebra was proposed for efficient computation in Refs.~\cite{linalg_PRC,linalg_JPG}.

In the present work, we test several quadratures for $\mathbb{S}^2$ (two-dimensional sphere surface in three-dimensional space) and $SO(3)$ using shell-model interactions and discuss their accuracy in the projection method. This paper is organized as follows: in Sect.~\ref{sec:jproj} the angular-momentum projector is defined and various types of quadratures are introduced. Numerical tests were done to discuss their accuracy in Sect.~\ref{sec:benchmark}. Sect.~\ref{sec:summary} is devoted to the summary.

\section{Angular-momentum projection}
\label{sec:jproj}
The angular-momentum projector is defined as
\begin{eqnarray}
  \hat{P}^J_{MK}
  &=& \frac{2J+1}{8\pi^2}\int d\Omega
  D^{J*}_{MK}(\Omega) \hat{R}(\Omega)
 \label{eq:jproj}
\end{eqnarray}
where $\Omega=(\alpha,\beta,\gamma)$ is the Euler angles and its integral is
\begin{eqnarray}
  \int d\Omega
  &=& \int_0^{2\pi} d\alpha \int_0^\pi \sin\beta d\beta \int_0^{2\pi} d\gamma .
 \label{eq:dOmega}
\end{eqnarray}
$\hat{R}(\Omega)$ is the rotation operator  as
\begin{equation}
    \hat{R}(\Omega)= e^{i\hat{J}_z\alpha}e^{i\hat{J}_y\beta}e^{i\hat{J}_z\gamma}.
\end{equation}
$D^J_{MK}$ is the Wigner $D$-function and is defined as 
\begin{eqnarray}
  D^J_{MK}(\Omega) 
  &=& e^{iM\alpha} d^J_{MK}(\beta)  e^{iK\gamma}
\end{eqnarray}
where $d^J_{MK}$ is the Wigner small $d$-function \cite{ringschuck}. 
% ”polynomial”の定義次第ではJ次の多項式とみなせる Ref.\cite{grafpotts} (角田)
In general, the range of the integral of $\gamma$ in Eq. (\ref{eq:jproj}) should be $[0,4\pi]$ due to the bipartite structure of $SU(2)$. In most practical applications this range can be reduced to $[0,2\pi]$ since we usually use a Slater determinant or a  number-parity-conserved quasi-particle vacuum, which does not mix integer-spin and half-integer-spin components  \cite{bally_proj}.

In most beyond-mean-field methods with the angular-momentum projection, the wave function and its energy $E$ are evaluated by solving the generalized eigenvalue problem
\begin{eqnarray}
  \sum_{jK} H^{(J)}_{iM,jK} g_{jK}
  &=& E \sum_{jK} N^{(J)}_{iM,jK} g_{jK}
 \label{eq:geneig} 
\end{eqnarray}
\begin{eqnarray}
  H^{(J)}_{iM,jK} 
  &=&
  \langle \phi_i | \hat{H} \hat{P}^J_{MK} | \phi_j \rangle 
  \label{eq:hmat}\\
  &=& \frac{2J+1}{8\pi^2}\int d\Omega
  D^{J*}_{MK}(\Omega) \langle \phi_i | \hat{H} \hat{R}(\Omega) | \phi_j \rangle
    \nonumber\\
  N^{(J)}_{iM,jK} 
  &=&
  \langle \phi_i | \hat{P}^J_{MK} | \phi_j \rangle
  \label{eq:nmat}\\
  &=& \frac{2J+1}{8\pi^2}\int d\Omega
  D^{J*}_{MK}(\Omega) \langle \phi_i | \hat{R}(\Omega) | \phi_j \rangle
    \nonumber
\end{eqnarray}
where $| \phi_i \rangle $ is a many-body basis state such as a Slater determinant or a number-projected quasi-particle vacuum. Thus, it is required to calculate the Hamiltonian and norm matrix elements in Eqs. (\ref{eq:hmat}) and (\ref{eq:nmat}) 
%($H^{(IMK)}_{ij}$ and $N^{(IMK)}_{ij}$) 
with high accuracy.
Usually, the integration in Eqs. (\ref{eq:hmat}) and (\ref{eq:nmat}) is performed numerically using Gaussian quadratures. 
In the present work, we aim at revealing mathematical conditions for the quadrature and discuss its properties for efficient computation.

This section is organized as follows: the mathematical aspect of quadrature in angular-momentum projection is discussed in Subsect.~\ref{sec:exact}.  We introduce several quadratures: the trapezoidal quadrature, the Gauss-Legendre quadrature in Subsect.~\ref{sec:gl},  the Lebedev quadrature  in Subsect.~\ref{sec:lebedev},  the $\mathbb{S}^2$ spherical design  in Subsect.~\ref{sec:sphdsn}, and  $SO(3)$ quadratures in Subsect.~\ref{sec:s3sd}, respectively.

% For numerical test, we use a Slater determinant provided by the Hartree-Fock calculations with the variation after angular-momentum projection and discuss the numerical accuracy of the expectation values of the energy and angular-momentum operator $\hat{J}^2$.

\subsection{Exactness of quadrature in projection method}
\label{sec:exact}

In this subsection we briefly mention the definition of exactness of quadrature and clarify necessary conditions of a quadrature with which the angular-momentum projection is performed exactly.

In numerical calculations, the integral of a general function $f(x)$ is evaluated by 
\begin{equation}
    \int_a^b dx\, f(x) = \sum_{i=1}^N w_i f(x_i)
\end{equation}
where $x_i$ is a sampling point in the range $[a, b]$ and $w_i$ is its corresponding weight with $\sum_{i=1}^N w_n=b-a$. A rule to determine $(x_i, w_i)$ for efficient computation is called Gauss-type quadrature.  
On the other hand, equally-weighted summation is also used as
\begin{equation}
    \int_a^b dx\, f(x) = \sum_{i=1}^N \frac{b-a}N  f(x_i).
\end{equation}
This type of quadrature is called the Chebyshev type  \cite{graef_thesis}. 
A set of the points and weights is designed so that the quadrature is mathematically exact and the number of the points $N$ is taken as small as possible under the condition that the integrand is any polynomial of degree at most $t$. 

% $d$-sphereの説明を想定される読者層に分かりやすく
We introduce the degree of exactness of a quadrature rule on $d$-sphere $\mathbb{S}^d$ and the rotation group $SO(3)$ \cite{graf,grafpotts}. 
A $d$-sphere is a $d$-dimensional sphere surface in ($d+1$)-dimensional space. 
A quadrature rule on $\mathbb{S}^d$ has a degree of exactness $t$ if the integral of any spherical harmonics of degree at most $t$ is exact mathematically.
% This condition means that the quadrature enables us to integrate any spherical harmonics of degree at most $t$ exactly. 
A quadrature rule on $SO(3)$ has a degree of exactness $t$ if the integral of any Wigner $D$-function $D^J_{MK}$ is exact with $J$ being an integer and  $J\le t$.
% Jが整数であることを条件に入れる

Hereafter, we discuss the condition that the integral appeared in the angular-momentum projection is carried out exactly. An arbitrary wave function $|\phi\rangle$ is written as a linear combination of the normalized, angular-momentum-projected states $|I, K\rangle$, which are eigenstates of $\hat{J}^2$ and $\hat{J}_z$ with eigenvalues $I(I+1)$ and $K$,  as 
\begin{equation}
    |\phi\rangle = \sum_{I=0\,\textrm{or}\,\frac12}^{I_\textrm{max}}\sum_{K=-I}^I v_{IK}|I,K\rangle
\end{equation}
where $v_{IK}$ is a coefficient of the linear combination and $I_\textrm{max}$ is the maximum angular momentum contained in the wave function.
The rotation of $|I, K\rangle$ is represented as
\begin{equation}
  \hat{R}(\Omega)|I,K\rangle=\sum_{M=-I}^I D^I_{MK}(\Omega)|I,M\rangle.
\end{equation}
Then, the angular-momentum projected state is 
\begin{eqnarray}
 \hat{P}^J_{MK} |\phi \rangle &= & \frac{2J+1}{8\pi^2}\int d\Omega D^{J*}_{MK}(\Omega) 
 \label{eq:projwf}\\
 &\times&\sum_{I=0\,\textrm{or}\,\frac12}^{I_\textrm{max}}\sum_{M',K'=-I}^{I}v_{IK'} D^{I}_{M'K'}(\Omega) | I, M' \rangle, \nonumber
% && \hat{P}^J_{MK} |\phi \rangle \\
% &=& \frac{2J+1}{8\pi^2}\int d\Omega   D^{J*}_{MK}(\Omega)\sum_{J'K'M'} v_{J'K'} D^{J'}_{M'K'}(\Omega) | J', M' \rangle, 
% \nonumber
\end{eqnarray}
which should be equal to $v_{JK}|J, M\rangle$ due to the property of the projection operator.
Therefore, for any $v_{JK}$, the projection can be performed exactly if the orthogonality condition
\begin{equation}
    \delta_{JI}\delta_{MM'} \delta_{KK'}= \frac{2J+1}{8\pi^2}\int d\Omega   D^{J*}_{MK}(\Omega) D^{I}_{M'K'}(\Omega)
    \label{eq:ddproj}
\end{equation}
is satisfied by a quadrature numerically.
% The product of $D^{J*}_{MK}(\Omega)$ and $D^{J'}_{M'K'}(\Omega)$ 
This integrand
is given as 
\begin{eqnarray}
  & &D^{J*}_{MK}(\Omega)D^{I}_{M'K'}(\Omega)\nonumber\\
  &=&\sum_{J'=|J-I|}^{J+I}(-1)^{M-K}\langle J,-M,I,M'|J',M'-M\rangle\nonumber\\
  &\times&\langle J,-K,I,K'|J',K'-K\rangle D^{J'}_{M'-M,K'-K}(\Omega)
\end{eqnarray}
where $\langle J,M,I,M'|J',M''\rangle$ is the Clebsch-Gordan coefficient. 
% This equation satisfies the condition $J'\le J+I\le J+I_\textrm{max}$.
Thus, the integrand of Eq.~(\ref{eq:ddproj}) is proved to be a linear combination of the $D^{J'}_{M'-M,K'-K}(\Omega)$ with $J'\le J+I_\textrm{max}$. 
Eq.~(\ref{eq:projwf}) is represented as
\begin{eqnarray}
 \hat{P}^J_{MK} |\phi \rangle &=& \sum_{I,M',K',J'} c^{J,I,J'}_{M,M',K,K'} | I, M' \rangle \nonumber\\
 &\times& \int d\Omega D^{J'}_{M'-M,K'-K}(\Omega) , 
 \label{eq:projwfd}
\end{eqnarray}
where $c^{J,I,J'}_{M,M',K,K'}$ are coefficients depending on their indices and $v_{IK'}$, 
% and numerically integrated in practice. 
and independent of $\Omega$.
% The quadrature of the exactness $t=J+I_\textrm{max}$ is required for the exact integral.
As a consequence, the integral in  Eq.~(\ref{eq:projwf}) can be performed exactly by using a quadrature rule with a degree of exactness $t=J+I_\textrm{max}$. %for the projection of a wave function with angular momentum at most $J_\textrm{max}$ onto angular momentum $J$.
% $D^{J''}_{M'-M,K'-K}(\Omega)$ with $J''\le J+J'\le J+J_\textrm{max}$ are integrated for projection of a wave function with angular momentum at most $J_\textrm{max}$ onto angular momentum $J$.

The determination of sampling points and their weights for an efficient quadrature with the degree $t$ is non-trivial and even challenging for a multivariable function such as the rotation group $SO(3)$.
In the following subsections, we apply several quadrature rules to the projection method and discuss their efficiency.

\subsection{Trapezoidal and Gauss-Legendre quadratures}
\label{sec:gl}

The most straightforward way to compute the three-fold integral is to compute them individually. The matrix elements in Eqs. (\ref{eq:hmat}) and (\ref{eq:nmat}) are calculated by the three-times integrations individually.
Traditionally, the trapezoidal quadrature is used for the integral of $\alpha$ and $\gamma$, and the Gauss-Legendre quadrature is used for the integral of $\beta$ \cite{bally_proj}. In the present work, it is referred to as the T+GL+T method.
Hereafter, we discuss the necessary conditions for the exactness of the quadratures in the T+GL+T method.

Eq.~(\ref{eq:projwfd}) is represented as a product of three integrals as 
\begin{eqnarray}
 \hat{P}^J_{MK} |\phi \rangle &=& \sum_{I,M',K',J'} c^{J,I,J'}_{M,M',K,K'} | I, M' \rangle\nonumber\\
 &\times& \int_0^{2\pi} d\alpha\,e^{i(M'-M)\alpha} \nonumber\\
 &\times& \int_0^\pi \sin\beta d\beta\,d^{J'}_{M'-M,K'-K}(\beta) \nonumber\\
 &\times& \int_0^{2\pi} d\gamma\,e^{i(K'-K)\gamma}. \label{eq:projwftri}
\end{eqnarray}
At first, we discuss the integral of $\gamma$ in Eq.~(\ref{eq:projwftri}), which corresponds to the $K$ projection. 
It is numerically performed by the summation
with the trapezoidal rule as
\begin{equation}
    \int_0^{2\pi}d\gamma\,e^{i(K'-K)\gamma}
    =  \frac{2\pi}{N_z} \sum_{m=1}^{N_z}
    e^{i(K'-K)\frac{2\pi m}{N_z}}
    \label{eq:trapem}
\end{equation}
where $N_z$ is the number of points for the quadrature.
This trapezoidal rule is equivalent to 
the Fomenko formula \cite{fomenko,linalg_JPG}
and is quite efficient despite its simplicity
because of the periodicity of this function.
If we consider all $K$ for a specific $J$, the integral of Eq.~(\ref{eq:trapem}) is exact under the condition $J+I_\textrm{max} < N_z$ because $|K'-K| \le J+I_\textrm{max}$. 
Thus, the minimum number of the $N_z$ for exact quadrature is 
\begin{equation}
    N_z = J + I_{\textrm{max}} + 1 = t+1.
\end{equation}
In the same way, the integral about $\alpha$ of the $M$ projection is performed. 

On the other hand, the integral of $\beta$ is performed by the Gauss-Legendre quadrature efficiently. 
Note that the Gauss-Legendre quadrature should be applied
to the function of $\cos\beta$,
not $\beta$ itself \cite{bally_proj}, as
\begin{equation}
  \label{eq:jylg}
  \int_0^{\pi} d\beta \sin\beta = \int_{-1}^{1} d(\cos\beta)  
\end{equation}
for high accuracy.
Since the integral of $\gamma$ ($\alpha$) of Eq.~(\ref{eq:projwftri}) becomes zero in the case of $K\ne K'$ ($M\ne M'$),
we need to consider the integral of $\beta$ only for $K=K'$ and $M=M'$.
% and the value of Eq.~(\ref{eq:projwftri}) is independent of the third line of Eq.~(\ref{eq:projwftri}).
In this case, the integral of $\beta$ in Eq.~(\ref{eq:projwftri}) is represented as
\begin{equation}
  \int_0^\pi \sin\beta d\beta\,d^{J'}_{00}(\beta)
  = \int_{-1}^{1} d(\cos\beta) P_{J'}(\cos\beta)
\end{equation}
where $P_{J'}$ is the Legendre polynomial.
The degree of the polynomial is $t=J+I_\textrm{max}$ at most and the Gauss-Legendre quadrature is exact if the number of points is equal to or larger than
\begin{equation}
    N_y = \lceil (t+1)/2 \rceil
\end{equation}
where $\lceil (t+1)/2 \rceil$ denotes the minimum integer which is equal to or larger than $(t+1)/2$ \cite{mclaren}.
Thus, the number of points for the T+GL+T method is estimated as
\begin{equation}
    N_\textrm{T+GL+T}
    = N_z^2 N_y = (t+1)^2 \lceil (t+1)/2 \rceil
    = \frac12 t^3 + O(t^2).
\end{equation}

However, the sampling points of the T+GL+T method are not equally distributed on the SO(3) manifold: the points are dense around the poles of $\beta=0$ and $\beta=\pi$ while they are sparse around the equator, $\beta=\pi/2$. While this feature might be suitable for an axis-aligned wave function \cite{taniguchi_align}, general variation-after-projection methods provide us with a randomly directed wave function.
We expect a quadrature with points that are equally distributed on the manifold is more efficient.
In the next two subsections, we introduce two  quadratures for $(\alpha, \beta)$ whose sampling points are distributed almost equally on $\mathbb{S}^2$.

\subsection{Lebedev quadrature}
\label{sec:lebedev}

For the integral over the spherical surface $\mathbb{S}^2$, the Lebedev quadrature was proposed and known to be quite efficient \cite{lebedev1,lebedev2}.
In this section, we introduce a product of the Lebedev and trapezoidal quadratures for $SO(3)$.
The points of the Lebedev quadrature keep 
the symmetry of octahedral rotation and inversion, and are distributed almost equally on $\mathbb{S}^2$. 
The points and their weights are determined so that the quadrature is exact for any spherical harmonics with degree up to $t$. The sampling points and weights for up to 131st-degree polynomial are available. 
The Lebedev quadrature has been used in various fields, such as the orientation averaging of the nuclear magnetic resonance \cite{nmr-stevensson}.
It was also applied to the spin projection in quantum chemistry 
and shows better performance than the T+GL+T quadrature \cite{chem}.

\begin{figure}[htbp]
  \includegraphics[scale=0.4]{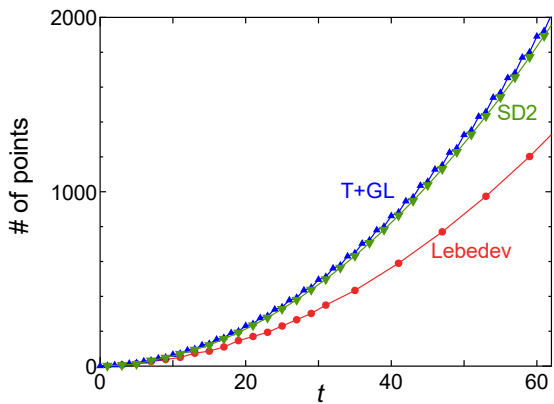}
  \caption{
    Number of points for the Lebedev quadrature (red circles),  trapezoidal+Gauss-Legendre (T+GL) quadrature (blue triangles), and the spherical design (SD2, green inverse triangles) to calculate the integral of a polynomial of at most $t$ degree exactly for the unit sphere surface $\mathbb{S}^2$ ($\alpha$ and $\beta$).
    For the T+GL method,  $(t+1)\lceil (t+1)/2 \rceil$ is shown.
  }
  \label{fig:quad-order}
\end{figure}

Figure~\ref{fig:quad-order} shows the necessary number of points of the Lebedev quadrature for $\mathbb{S}^2$ against the degree $t$. The number is given as \cite{lebedev1,lebedev2}  
\begin{equation}
    N_\textrm{Lebedev}\simeq \frac13 (t+1)^2 + 2
\end{equation} 
For comparison, we show the number of points for the trapezoidal and Gauss-Legendre quadratures for $\alpha$ and $\beta$, which is referred to as the T+GL method. The number of the points of the Lebedev quadrature is smaller than that of the T+GL method by a factor of 2/3.

For the integral of the projection operator, we adopt the Lebedev quadrature for  $(\alpha, \beta)$ and the trapezoidal quadrature for $\gamma$, which is referred to as the Lebedev+T method.
Since the numbers of points $N$ for quadrature with degree of exactness $t$ are $N \simeq \frac13 (t+1)^2+2$ for the Lebedev and $N=t+1$ for the trapezoidal quadrature, that for the Lebedev+T method is
\begin{equation}
    N_\textrm{Lebedev+T} \simeq (t+1) \left( \frac13 (t+1)^2+2\right).
\end{equation}
Similar to the discussion of the T+GL+T method, we consider the integral of $(\alpha, \beta)$ only for $K=K'$.
The integrand of Eq.~(\ref{eq:projwfd}) is
\begin{equation}
  D^{J'}_{M'-M,0}(\alpha,\beta,\gamma)=(-1)^{M'-M}\sqrt{\frac{4\pi}{2J'+1}}Y^{(J')}_{M'-M}(\beta,\alpha)
\end{equation}
%$D^J_{MK}$ with $K\ne0$ is exactly integrated as 0 by the trapezoidal quadrature.
%In case of $K=0$, 
%%式が正しいか要チェック
%\begin{equation}
%  D^J_{M0}(\alpha,\beta,\gamma)=\sqrt{\frac{4\pi}{2J+1}}Y^{(J)}_M(\beta,\alpha)
%\end{equation}
and spherical harmonics are exactly integrated by the Lebedev quadrature up to the degree $t$.
% K=0 以外では？？ <- T+GL+Tの追記部分の説明を参照

\begin{figure}[htbp]
  \includegraphics[scale=0.4]{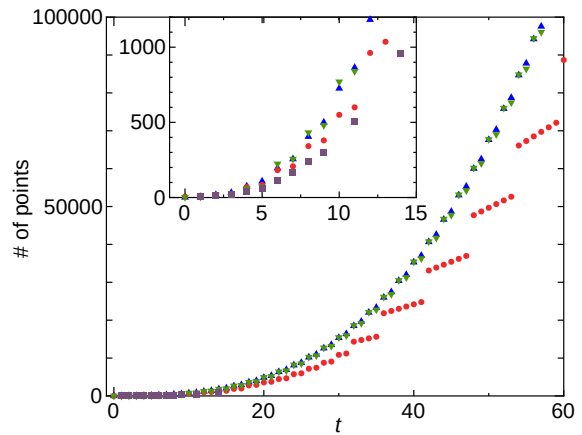}
  \caption{
    Number of points for the Lebedev+trapezoidal quadrature (red circles), the spherical design+trapezoidal quadrature (green inverse triangles),  the Gauss-Legendre+trapezoidal quadrature (blue triangles), and the Gauss-type $SO(3)$ quadrature (purple squares) to calculate the integral of the function of at most degree $t$ exactly for the Euler angles. 
    The inset shows a magnified view.
      }
  \label{fig:order-points}
\end{figure}

Figure \ref{fig:order-points} shows the number of points for the Lebedev method against the degree $t$. The discontinuous behavior of the Lebedev quadrature in Fig.~\ref{fig:order-points} is because only the points and degrees of the Lebedev quadrature of the discrete degree are available as shown in Fig.~\ref{fig:quad-order}.
Apparently, the Lebedev method outperforms the T+GL+T method a factor 2/3 at most.

\subsection{Spherical design on $\mathbb{S}^2$}
\label{sec:sphdsn}

A spherical design is a part of the combinatorial design theory in mathematics and has a wide variety of applications.
In this theory, the sampling points are equally distributed on the sphere surface $\mathbb{S}^2$ as far as possible \cite{spherical_design}.
The points of the spherical design are used as sampling points of a Chebyshev-type quadrature, namely equally weighted summation.
Ref.~\cite{beentjs} demonstrates that this quadrature shows competing performance with the Lebedev method for a certain type of integrand functions.

The number of points of the spherical design required for the exactness $t$ (spherical $t$-design) is estimated as \cite{spherical_design}
\begin{equation}
    N_\textrm{SD2} = \frac12 t^2 + \frac12 t + O(1)
\end{equation}
and is shown in Fig.~\ref{fig:quad-order}. While it shows similar behavior to the T+GL method, slight improvement is seen.  The spherical design shows lower performance than the Lebedev method since the spherical design is an equally weighted quadrature but in the Lebedev method the degree of freedom of weights is also used for high accuracy.

For the projection, we test a product of the spherical design for the integral of $(\alpha, \beta)$ and the trapezoidal quadrature for that of $\gamma$, which is referred to as the SD2+T method. In Fig. \ref{fig:order-points} the number of the SD2+T is shown in a similar tendency to that of the GL+T method.
The number of points is estimated as 
\begin{equation}
    N_\textrm{SD2+T} = \frac12 t^3 + t^2 + O(t)
\end{equation}
whose leading-order term agrees with that of the T+GL+T method.

\subsection{$SO(3)$ quadratures}
\label{sec:s3sd}

Here, we discuss a possible quadrature for $SO(3)$ in a single scheme.
The spherical design of $\mathbb{S}^3$ (three-dimensional sphere surface in four-dimensional space) can be transformed into the quadrature of Euler angles \cite{graf}. %because $SU(2)$ is a bipartite double covering of $SO(3)$. 
Ref.~\cite{graf} proves that the $t$-degree quadrature of $SO(3)$ is equivalent to the $(2t+1)$-degree quadrature of  $\mathbb{S}^3$ having a point symmetry. 
We have tested some cases using such Chebyshev-type $SO(3)$ quadrature based on the spherical design $\mathbb{S}^3$ \cite{spherical_design}, but it does not outperform the Lebedev+T quadrature. 
Its number of points as a function of degree shows similar behavior to the case of the Gauss-Legendre case. 
% It is not shown in Fig.~\ref{fig:order-points} for simplicity.
 
The Gauss-type $SO(3)$ quadrature was suggested in Ref.~\cite{graef_thesis}. The quadrature rules are taken to be invariant under the tetrahedral group, octahedral group, or the icosahedral group symmetries.
Its points (Euler angles) and their weights are determined so that a quadrature for degree $t$ is exact on the $SO(3)$ manifold by numerical optimization \cite{graef_thesis}.
Its number of points is shown as the purple squares in the inset of Fig.~\ref{fig:order-points}. This quadrature shows the smallest number of the points at a certain degree and then seems promising.
However, to the best of our knowledge, high-degree Gauss-type $SO(3)$ quadrature is not available  (only the 14th-degree set with 960 points or lower ones are available), which restricts its practical application to a small problem.

\section{Numerical test}
\label{sec:benchmark}

For the benchmark test to discuss the accuracy of the projection, we prepare a symmetry-broken Slater determinant $|\phi\rangle$ given by the projected Hartree-Fock calculation (JHF) with the variation after angular-momentum (and parity, if necessary) projection. 
The JHF wave function is calculated by our MCSM code \cite{phys_scr_mcsm} since the first basis state of the MCSM wave function corresponds to the JHF solution.
The expectation values of the shell-model Hamiltonian and angular-momentum are calculated utilizing three quadratures introduced in the previous section: the T+GL+T, the Lebedev+T, and the SD2+T methods. By changing the degree $t$, namely the number of the sampling points,  the deviation from the exact value is discussed as an error of the quadratures.
The whole computations were performed in double precision.

\subsection{$^{57}$Fe with the $pf$-shell model space}

As a first numerical test, we take $^{57}$Fe with the $pf$-shell model space and the GXPF1A interaction \cite{gxpf1a}.
\begin{figure}[htbp]
  \includegraphics[scale=0.4]{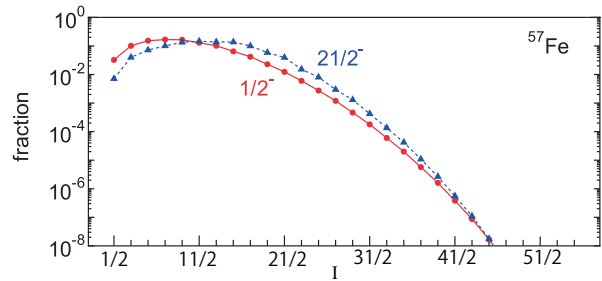}
  \caption{
   $I$-projected components, $f_I$, of the JHF wave functions of $^{57}$Fe.
   The wave functions are given by the JHF calculations with $J^\pi=1/2^-$ and $21/2^-$.
  }
  \label{fig:frac-fe57-j1n}
\end{figure}
Figure \ref{fig:frac-fe57-j1n} shows the shell-model wave function components, $f_{I}=\sum_K |v_{IK}|^2 = \sum_K \langle\phi| P^I_{KK} |\phi\rangle $, of the intrinsic wave function of the $J^\pi=1/2^-$ and $J^\pi=21/2^-$ states of $^{57}$Fe given by the JHF calculation.
With increasing $I$, its fraction decreases exponentially and becomes negligible. A similar tendency is seen in the case of the cranked Hartree-Fock calculation \cite{linalg_PRC}.

Since this model space contains $I_\textrm{max}=55/2$, quadratures for $t=28$ and $t=38$ is required  in principle for the $J^\pi=1/2^-$ and $21/2^-$ states.
However, since the JHF wave function does not contain higher-$I$ component, $I_\textrm{max}$ is reduced effectively.
According to Fig.~\ref{fig:frac-fe57-j1n}, the effective maximum number of the $I$ component with the threshold of $f_I \geq 10^{-5}$ is $I_\textrm{max} = 35/2$ for $J^\pi=1/2^-$ and $I_\textrm{max} = 37/2$ for $J^\pi=21/2^-$.
Therefore, the quadrature with degree $t=18$ and $t=29$ is practically required for $J^\pi=1/2^-$ and $J^\pi=21/2^-$ states, respectively.

\begin{figure}[htbp]
  \includegraphics[scale=0.4]{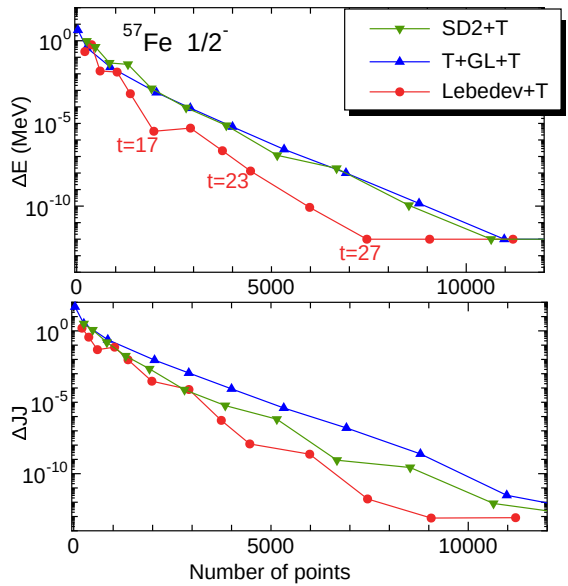}
  \caption{
   Error of the Hamiltonian and $J^2$ expectation values, $\Delta E =|E - E_\textrm{exact}|$ and $\Delta J^2 =|\langle \hat{J}^2 \rangle -J(J+1)|$,
   of the $1/2^-$ state of $^{57}$Fe against the number of points. The wave function is given by the JHF calculation.
  }
  \label{fig:Fe57-j1-diff}
\end{figure}

Figure \ref{fig:Fe57-j1-diff} shows the errors of the expectation values of $\hat{H}$ and $\hat{J}^2$ of the $J^\pi=1/2^-$ wave function with changing the degree $t$.
The error is defined as  $\Delta E =|E - E_\textrm{exact}|$ and $\Delta J^2 =|\langle \hat{J}^2 \rangle -J(J+1)|$, where $E$ is given by Eq.~(\ref{eq:geneig}).
$ E_\textrm{exact}$ is calculated by quadrature with sufficiently large number of points.
As the number increases the accuracy of quadratures is improved exponentially.
The Lebedev+T method shows the best performance in comparison with the T+GL+T and SD2+T methods. To obtain the same precision, the number of points for the Lebedev+T method is two-thirds of the T+GL+T method, which is consistent to the discussion in Subsect.~\ref{sec:lebedev}. While the SD2+T method shows similar performance to the T+GL+T method, the SD2+T method slightly outperforms the T+GL+T method.
The error of the energy with the quadrature $t=18$, which is determined by the fraction of the wave function, is expected to be around $10^{-5}$ MeV and small enough for practical usage. At $t=27$ the precision reaches the order of machine epsilon.

\begin{figure}[htbp]
  \includegraphics[scale=0.4]{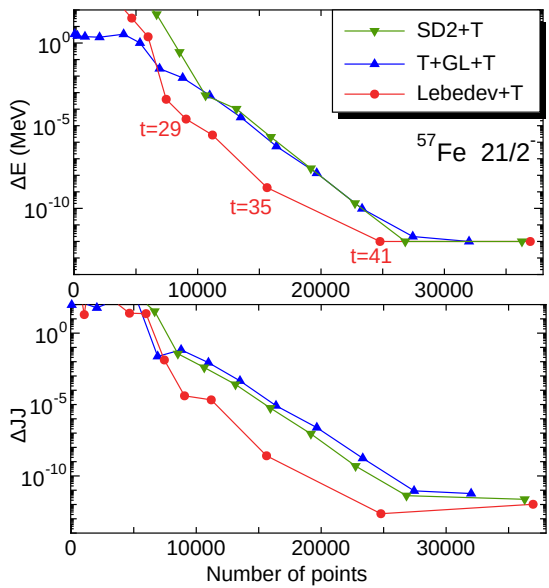}
  \caption{
   Errors of the Hamiltonian and $J^2$ expectation values of the $21/2^-$ state of $^{57}$Fe against the number of the quadrature points. See caption of Fig.~\ref{fig:Fe57-j1-diff} for details.
  }
  \label{fig:Fe57-j21-diff}
\end{figure}

Figure \ref{fig:Fe57-j21-diff} shows the errors of the expectation values of $\hat{H}$ and $\hat{J}^2$ in case of the $J^\pi=21/2^-$ wave function.
Similar tendency is shown to Fig. \ref{fig:Fe57-j1-diff}, except that the number of the points is increased since the degree $t$ is large for $J^\pi=21/2^-$.
The energy error with the quadrature $t=29$, which is determined by the fraction of the wave function, is around $10^{-5}$ MeV and small enough for practical usage.

\subsection{$^{10}$Be with no-core shell-model approach}

As a benchmark test for the no-core shell-model approach, we adopt the ground state of $^{10}$Be with the JISP16 interaction \cite{jisp16,tabe_4n}. The model space is taken as the 5 major shells ($s, p, sd, pf,$ and $sdg$ shells) and the harmonic-oscillator energy is  $\hbar\omega=25$ MeV.
The intrinsic wave function $|\phi\rangle$ is provided by the JHF calculation and the parity projection is also performed before variation.

\begin{figure}[htbp]
  \includegraphics[scale=0.4]{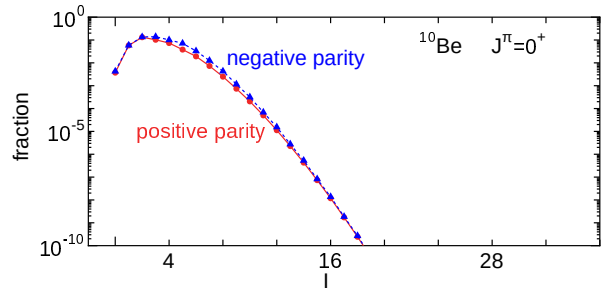}
  \caption{
   $I^\pi$-projected components of the JHF wave functions of the $J^\pi=0^+$ state of $^{10}$Be. The red circles (blue triangles) show positive-parity (negative-parity) projected components. 
  }
  \label{fig:frac-be10-j0p}
\end{figure}
Figure \ref{fig:frac-be10-j0p} shows the $I$-component of the $^{10}$Be ground-state wave function, $f_{I^\pi} = \sum_K \langle \phi|\hat{P}^I_{KK} \hat{P}^\pi|\phi \rangle$. $\hat{P}^\pi$ is the parity-projection operator and does not affect the accuracy of the numerical result.
The peak of the fraction is $I^\pi=2^+$ and it decreases exponentially as a function of $I$ resulting $I_\textrm{max}= 12$ with the threshold $f_I\geq 10^{-5}$ while the model-space restriction provides $I_\textrm{max}=35$. 

\begin{figure}[htbp]
  \includegraphics[scale=0.4]{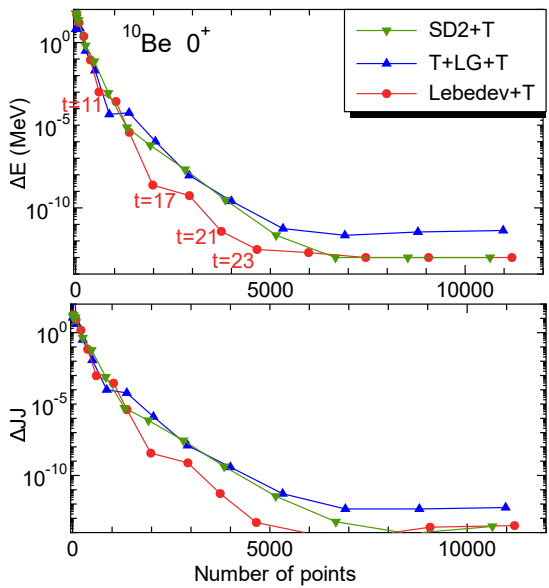}
  \caption{
 Errors of the expectation values of the Hamiltonian and $J^2$ of the ground state wave function of $^{10}$Be given by the JHF calculation.
 See caption of Fig.~\ref{fig:Fe57-j1-diff} for details.
  }
  \label{fig:be10-j0-diff}
\end{figure}
Figure \ref{fig:be10-j0-diff} shows the errors given by the angular-momentum projected expectation values of the Hamiltonian and angular-momentum operators.
With $I_\textrm{max}= 12$ and $t=12$, the error is around $10^{-3}$ MeV and rather larger than the case of $^{57}$Fe. However, the error of the quadrature with $t=19$ is $10^{-8}$ MeV, which is small enough for practical applications.

\subsection{$SO(3)$ quadrature}

In the previous subsections, we discussed the combination of the quadratures for ($\alpha$, $\beta$) and for $\gamma$. In this subsection, we discuss a quadrature for the rotation group $SO(3)$. We test a Gauss-type $SO(3)$ quadrature which is given by putatively optimal quadrature rules on the rotation Group $SO(3)$ and was proposed in \cite{graef_thesis}.

To discuss its performance, we performed a benchmark test utilizing the $J^\pi=4^+$ state of $^{22}$Ne with the $sd$-shell model space and the USD interaction \cite{usd}. 
The maximum angular momentum allowed in the model space is $I_\textrm{max}=10$.
Since the wave function has a certain fraction of the $I=10$ component, a quadrature with the degree $t=14$ is required. We numerically confirmed that at least a 14-degree quadrature is required to obtain a meaningful result for the four types of quadratures shown in Tab.~\ref{tab:14degree}.

\begin{table}[htbp]
    \centering
    \begin{tabular}{c|c}
      Method   & Number of points \\
      \hline\hline
        T+GL+T           & 1800 \\
        Lebedev+T        & 1290 \\ % Lebedev rank 15 * T 15
        SD2+T            & 1800 \\
        Gauss-type SO(3) &  960 \\
        Efficiency=1     & 1124
    \end{tabular}
    \caption{Number of points of quadratures for a degree of exactness $t=14$.}
    \label{tab:14degree}
\end{table}
Table \ref{tab:14degree} shows the numbers of the points for 14-degree quadratures.
The Lebedev+T method outperforms the conventional T+GL+T method around 30\%, 
the SD2+T method shows the same number of points with the T+GL+T method.
The Gauss-type $SO(3)$ quadrature outperforms them and seems to be promising.

Here, we briefly discuss how many points are required in principle for $SO(3)$ quadrature \cite{mclaren,graef_thesis}. For the quadrature with degree $t$, its points and weights are determined so that any function which is a linear combination of the Wigner $D$-functions up to rank $t$ is integrated exactly. 
The degree of freedom of the $D$-function with rank $J$ is $(2J+1)^2$, and hence the total number of the degree of freedom up to degree $t$ is $\sum_{J=0}^t (2J+1)^2$. Since each sampling point has four variables ($\alpha_i,\beta_i,\gamma_i$, and $w_i$), the minimum number of points required is estimated as 
\begin{equation}
    N_\textrm{eff} = \frac14 \sum_{J=0}^t (2J+1)^2 = \frac13 t^3+t^2 + \frac{11}{12}t+\frac14 .
\end{equation}
$N_\textrm{eff}$ of $t=14$ is shown as ``Efficiency=1'' in Table \ref{tab:14degree}. It is close to the number of the Lebedev+T method. In this case, the number of the Gauss-type $SO(3)$ quadrature is smaller than $N_\textrm{eff}$, which can be achieved by employing invariant theory, in which  quadrature rules are invariant under certain group symmetries  \cite{mclaren,graef_thesis}.
The number of points of the Lebedev+T quadrature becomes close to $N_\textrm{eff}$ asymptotically because the coefficient of the leading order is the same. Therefore we expect that it is sufficiently efficient in high-degree cases.
Nevertheless, further investigation of the Gauss-type $SO(3)$ quadrature is expected.

\section{Summary}
\label{sec:summary}

The angular-momentum projection requires a three-fold integral of Euler angles, which causes a burden of computations.
We discussed the necessary number of points of quadratures to perform the angular-momentum projection. Since it is approximately proportional to the computation time, the efficient computation of the integral has a large impact on various nuclear-structure studies.
We proved that the quadrature of the degree of exactness $t=J+I_\textrm{max}$ is required for the exact numerations where $J$ is the angular momentum of the projection and $I_\textrm{max}$ is the maximum angular momentum contained in the wave function before projection.

For numerical tests, we adopted mainly three methods: the T+GL+T, the Lebedev+T, and the SD2+T quadratures. 
With $t=J+I_\textrm{max}$, the conventional T+GL+T quadrature becomes mathematically exact if the number of points are taken as $N_z\geq t+1$ and $N_y\geq \lceil (t+1)/2 \rceil$.
For the Lebedev quadrature and the spherical design, the data sets of degree $t$ give us the exact quadrature. 
The number of sampling points is $N\sim \frac 13 t^3$ for the Lebedev+T quadrature and $N\sim \frac 12 t^3$ for the conventional T+GL+T quadrature asymptotically. Thus, the number of the points and namely the computation time is reduced by the factor 2/3 by introducing the Lebedev quadrature.
The SD2+T quadrature shows slightly better behavior to the T+GL+T case.
In addition, the Gauss-type $SO(3)$ quadrature was also discussed showing a promising result.
This discussions are also applicable to the isospin projection.

If we apply a symmetry restriction to the wave function to reduce the number of points \cite{enami}, the quadrature should have the same symmetry. 
It is desired to develop an efficient Gaussian $SO(3)$ quadrature having appropriate symmetries for higher degrees.

\section*{Acknowledgment}

The authors acknowledge Keigo Nitadori for stimulating suggestion, Hiroshi Murakami for useful information, Samuel Gr\"{a}f, Takahiro Mizusaki and Takaharu Otsuka for valuable comments.

This research was supported 
by ``Program for Promoting Researches on the Supercomputer Fugaku''
(JPMXP1020200105) and JICFuS, 
and KAKENHI grant (17K05433, 20K03981).
This research used computational resources of the supercomputer Fugaku
(hp220174, hp210165) at the RIKEN Center for Computational Science,  
the Oakforest-PACS supercomputer (xg18i035) for the MCRP program at Center for Computational Sciences, University of Tsukuba, 
and the Oakbridge-CX supercomputer.

\end{document}